\documentclass[aps,prx,twocolumn,groupedaddress,amsmath,epsfig,,eqsecnum,amssymb]{revtex4-1}
\usepackage{makecell}
\usepackage{bbm}
\usepackage{bm}
\usepackage{mathrsfs}
\usepackage{subfigure}
\usepackage{multirow} 
\usepackage{bigstrut} %

\usepackage{amsfonts}
\usepackage{color}
\usepackage{graphicx}

     \usepackage{hyperref} 
 \hypersetup{
 colorlinks=true,
 linkcolor=blue,
 urlcolor=blue,
 }

\newcommand{\dbar}{d\hspace*{-0.08em}\bar{}\hspace*{0.1em}}
\newcommand{\LFP}{{\mathcal{L}}}

\newcommand{\Pv}{\boldsymbol P}
\newcommand{\Hv}{\boldsymbol H}
\newcommand{\mv}{{\boldsymbol m}}

\newcommand{\fv}{{\boldsymbol f}}

\newcommand{\be}{\begin{equation}}
\newcommand{\ee}{\end{equation}}
\newcommand{\ba}{\begin{eqnarray}}
\newcommand{\ea}{\end{eqnarray}}

\begin{document}
\title{ Stochastic Thermodynamics of  Micromagnetics with Spin Torque}
\author{Mingnan Ding$^1$}
\author{Jun Wu$^1$}
\author{Xiangjun Xing$^{1,2,3}$}
\email{xxing@sjtu.edu.cn}
\address{$^1$Wilczek Quantum Center, School of Physics and Astronomy, Shanghai Jiao Tong University, Shanghai 200240, China \\
$^2$T.D. Lee Institute, Shanghai Jiao Tong University, Shanghai 200240, China\\
$^3$Shanghai Research Center for Quantum Sciences, Shanghai 201315, China}
\date{\today} 
	
	
\begin{abstract}
In this work, we study the stochastic dynamics of micro-magnetics interacting with a spin-current torque.  We extend the previously constructed stochastic Landau-Lifshitz equation to the case with spin-current torque, and verify the conditions of detailed balance.  Then we construct various thermodynamics quantities such as work and heat, and prove the second law of thermodynamics.  Due to the existence of spin-torque and the asymmetry of the kinetic matrix, a novel effect of {\em  entropy pumping} shows up.  As a consequence, the system may behave as a heat engine which constantly transforms heat into magnetic work.  Finally, we  {derive} a fluctuation theorem for the joint probability density function of the pumped entropy and the total work, and  {verify} it using numerical simulations. 

\end{abstract}
\maketitle 

\section{Introduction} 
\label{sec:intro}
In the previous work~\cite{LLG-1}, we studied the stochastic thermodynamics of a micro-magnet coupled to a magnetic field.   With the magnetic field fixed, such a system converges to a thermodynamic equilibrium state that obeys detailed balance and has no entropy production.  The probability current at equilibrium is however non-vanishing, due to the existence of so-called reversible current which does not contribute to dissipation. 

In the present work, we consider a micro-magnet driven by a non-conservative torque called {\em spin-current torque} or {\em spin-torque}, which arises due to interaction with a spin-polarized current.   Although interaction between  spin-current and magnetization is of quantum origin~\cite{Stiles2006}, the resulting dynamics of the magnetization can be still described by the classical  {Landau-Lifshitz equation} augmented by a new  term~\cite{Slonczewski1996,Berger1996,Brataas-Nature-materials-2012}.  In this work, we shall incorporate stochasticity into this dynamics and study its stochastic thermodynamics. We shall find a novel effect called {\em entropy pumping}, which we argue represents the exchange of entropy between the system and the spin-current that does not involve dissipation.  The effect of entropy pumping was firstly discovered by Kim and Qian~\cite{Kim2004,Kim2007} some time ago in Hamiltonian system driven by velocity dependent forces. 

This work is organized as follows.  In Sec.~\ref{sec:LLGS} we derive the stochastic  Landau-Lifshitz-Slonczewski ({\em sLLS}) equation, which describes the Langevin dynamics of a micro-magnet coupled both to a magnetic field and to a spin-torque, and discuss the conditions of local detailed balance.  In Sec.~\ref{sec:ST} we develop the theory of stochastic thermodynamics.  We establish the first and second laws of thermodynamics, and discuss the effect of entropy pumping.  In Sec.~\ref{sec:FT} we {  derive} the fluctuation theorem for the joint distribution of work and pumped entropy, and verify it using numerical simulation.  Finally in Sec.~\ref{sec:conclusion} we draw the concluding remarks.


\section{The Stochastic Landau-Lifshitz-Slonczewski equation}
\label{sec:LLGS}


\label{sec:torque}
As shown in Eq.~(2.12) of the preceding work~\cite{LLG-1}, a magnetic moment coupled to a magnetic field $\Hv$ evolves according to the following {\em stochastic Landau-Lifshitz (sLL)} equation:
\ba
{d \mv} =&  - & \gamma_0 \, \mv \times \Hv \, dt
- \eta \, \mv \times (\mv \times \Hv)\, dt
\nonumber\\
&-& 2 T \eta \,\mv \,dt 
+ {\sqrt{2 \eta T}} \mv \times d{\boldsymbol W}, 
\label{Langevin-def-m-vec}
\ea
where $\gamma_0$ and $\eta$ are respectively the gyromagnetic ratio and the damping coefficient, $d{\boldsymbol W}$ is vector-valued Wiener noise, whereas the product $ \mv \times d{\boldsymbol W}$ is defined in Ito's sense.  This equation can be put into the covariant form:
\ba
d m^i  + \left(\, L^{ij} (\partial_j U  )
- \partial_j L^{ij}\,\right) dt
=  b^{i \alpha} d W_{\alpha}(t) ,  \quad \quad
\label{Langevin-def-m-nonc}
\ea
where $U$ is  {\em the generalized potential}
\begin{subequations}
\label{LQBU-LLG}
\ba
 U(\mv, \Hv) &=& - \beta \mv \cdot \Hv - \beta F (\Hv),
 \label{LQBU-LLG-U}
\ea
whereas $F (\Hv)$ is the {\em equilibrium free energy}:
\ba
F  (\Hv) 
&=& - T \log \frac{4 \pi m\, \sinh \beta |\Hv| m}{\beta |\Hv|}. 
\label{F-lambda-def}
\ea
$L^{ij}$ is {\em the kinetic matrix} which can be decomposed into a symmetric part $B^{ij}$
 and an antisymmetric part $Q^{ij}$:
 \ba
 L^{ij}(\mv) &=& B^{ij} + Q^{ij}, \\
Q^{ij}(\mv)  &=&   T \gamma_0 \epsilon^{ijk} m_k,
  \label{LQBU-LLG-Q}\\
 B^{ij} (\mv)&=&  T \eta \left( m^2  \delta^{ij} - m^i m^j \right), 
  \label{LQBU-LLG-B}
\ea
whereas the matrix $b^{i \alpha}$ is given by
\ba
b^{i \alpha}(\mv) &=& 
\frac{\sqrt{2 \eta T}}{m} (m^2 \delta^{i\alpha}  - m^i m^\alpha). 
\label{LQBU-LLG-b}
\ea
\end{subequations}
If the magnetic field is fixed, the system converges to the following equilibrium state:
\ba
p^{\rm eq}(\mv, \Hv)
 = e^{- U(\mv, \Hv)}. 
 \label{p_eq-U-H}
\ea


The effect of {\em spin torque} may be taken into account by making the following replacement in Eq.~(\ref{Langevin-def-m-vec})~\cite{Stiles2006,Aron2014}:
 \ba
\Hv \rightarrow \Hv + \fv 
 = \Hv  +  \mv \times \Pv,
 \label{Hv-Hv-f-LLG}
\ea 
where the vector $\Pv$ is parallel to the spin-polarization of the current, which in general may depend on many details of the spin-current, as well as on the magnetization $\mv$ itself. We shall make the simplifying assumption that $\Pv$ is independent of $\mv$.  Below we shall call $\Pv$ {\em the spin current vector}, for simplicity.  Hence $\mv \times \Pv$ may be understood as the {\em nonconservative} effective magnetic field induced by the spin-polarized current.   It is nonconservative because if cannot be expressed as the gradient of certain potential function.   The resulting equation shall be called {\em stochastic Landau-Lifshitz-Slonczewski (sLLS) equation} :
\ba
{d \mv} =& - & \gamma_0 \mv \times ( \Hv + \mv \times \Pv) \, dt
\nonumber\\
&-&  \eta \, \mv \times (\mv \times ( \Hv + \mv \times \Pv))\, dt
\nonumber\\
&-& 2 T \eta \,\mv dt 
+ {\sqrt{2 \eta T}} \mv \times d{\boldsymbol W}.  
\label{Langevin-def-m-vec-NC}
\ea
This equation reduces to Eq.~(2.12) of Ref.~\cite{LLG-1} in the limit of vanishing spin current $\Pv \rightarrow 0$.  It is also equivalent to Eq.~(15) of Ref.~\cite{Aron2014} and Eq.~(5) of Ref.~\cite{Utsumi2015}, which were obtained by incorporating stochasticity and spin-torque into the stochastic Landau-Lifshitz-Gilbert equation. Equation (\ref{Langevin-def-m-vec-NC}) is more convenient for our discussion of stochastic thermodynamics. 


The sLLS equation (\ref{Langevin-def-m-vec-NC}) is a special case of the covariant Ito-Langevin equation driven by non-conservative forces~\cite{sto-therm-NC}:
\ba
d m^i  + \left(\, L^{ij} (\partial_j U - \varphi_j )
- \partial_j L^{ij}\,\right) dt
=  b^{i \alpha} d W_{\alpha}(t),
\quad \quad
\label{Langevin-def-m-nonc}
\ea
where ${\bm \varphi} $ is the rescaled non-conservative force:
\be
{\bm \varphi} \equiv \beta \, \mv \times \Pv .
\label{varphi-def}
\ee

Using  Eq.~(\ref{Langevin-def-m-vec-NC}) and Ito's rule:
\be
dW_\alpha dW_\beta = \delta_{\alpha \beta} dt,
\ee
we find the magnitude of $\mv$ is conserved by Eq.~(\ref{Langevin-def-m-vec-NC}):
\ba
d m^2 &=& d ( \mv \cdot \mv )
= 2 \mv \cdot d \mv + d\mv \cdot d \mv 
\nonumber\\
&=&- 4 T \eta \, m^2 dt + 4 T \eta \, m^2 dt  = 0. 
\ea

 
The Fokker-Planck equation (FPE) corresponding to the Langevin equation (\ref{Langevin-def-m-nonc}) is
\ba
\partial_t \, p(\mv,t) 
= \LFP  \, p(\mv,t) ,
\label{FPE-1} 
\ea
where the Fokker-Planck operator is derived in Ref.~\cite{sto-therm-NC}:
\be
\LFP \equiv \partial_i L^{ij} ( \partial_j + (\partial_j U ) - \varphi_j ).
\label{FPO-def}
\ee
Using the expressions for $L^{ij}, U$ and $\varphi_j$, we may  rewritte Eq.~(\ref{FPO-def}) into the following  form:
\begin{subequations}
\ba
\mathcal L &=& \mathcal L_0 + \delta \mathcal L, 
\label{FPO-LLG-NC}\\
\mathcal L_0   &=& T\eta (m^2 \nabla^2 - m^i m^j
 \partial_i \partial_j - 2 \mv \cdot \nabla )
\nonumber\\
&+& \eta (- m^2 \Hv \cdot \nabla + (\mv \cdot \Hv) \mv \cdot \nabla
+ 2 \mv \cdot \Hv)
\nonumber\\
&+& \gamma_0 (\mv \times \Hv) \cdot \nabla,
\label{FPO-LLG}\\
\delta \mathcal L &=&
\gamma_0 (- m^2 \Pv \cdot \nabla + (\mv \cdot \Pv) \mv \cdot \nabla
+ 2 \mv \cdot \Pv)
\nonumber\\
&-& \eta m^2 (\mv \times \Pv) \cdot \nabla,
\ea
\label{FPO-torque}
\end{subequations}
where $\mathcal L_0$ is the  Fokker-Planck operator  in the absence of spin torque, as given by Eq.~(2.15) of I, and $\delta \mathcal L$ is due to the spin torque.  The derivation is tedious but straightforward. 



The FPE (\ref{FPE-1}) may be rewritten into the following form of probability conservation:
\ba
\partial_t p = \partial_k j^k, 
\label{FPE-j}
\ea
where $j^i$ is the probability current defined as:
\be
j^k = -  L^{kj} (\partial_j + \partial_j U - \varphi_j ) p
 + \partial_j (Q^{kj} p). 
\label{current}
\ee
We may decompose the current into a {\em reversible current}  $j^i_{\rm R}$ and an {\em irreversible current} $j^i_{\rm IR}$~\cite{sto-therm-NC}:
\begin{subequations}
\ba
j^k_{\rm R} &=& -  Q^{kj} (\partial_j + \partial_j U - \varphi_j ) p 
+ \partial_j (Q^{kj} p),
\label{j_R-def}
\\
j^k_{\rm IR} &=& -  B^{kj} (\partial_j + \partial_j U - \varphi_j ) p. 
\label{j_IR-def}
\ea
\label{j_R-IR-def}
\end{subequations}
 {These names are pertinent because, as we will show below in Eq.~(\ref{dS^tot-NC-j-R}), the irreversible current $j^k_{\rm IR}$  but not the reversible current  $j^k_{\rm R}$ contributes to the entropy production.   Invoking Eqs.~(\ref{LQBU-LLG-Q}) and (\ref{LQBU-LLG-B}), we see that the reversible current $j^k_{\rm R}$ is proportional to the gyromagnetic ratio $\gamma_0$, whereas the irreversible current $j^k_{\rm IR}$ is proportional to the damping coefficient $\eta$.  }
We note that in the absence of spin torque, Eq.~(\ref{current}) reduces to Eq.~(2.20) of Ref.~\cite{LLG-1}, and Eqs.~(\ref{j_R-IR-def}) reduce to Eqs.~(2.21) of Ref.~\cite{LLG-1}.


\subsection{Detailed balance} 
For micromagnetic systems without spin torque, the conditions of detailed balance were discussed in Sec.~IID of Ref.~\cite{LLG-1}. These conditions guarantee that the steady state can be understood as a thermodynamic equilibrium.  In the presence of spin torque, these conditions should be properly generalized, and should be called {\em the conditions of local detailed balance}.  The conditions of local detailed balance for the covariant Langevin equation (\ref{Langevin-def-m-nonc})  driven by non-conservative forces were given in Eqs.~(3.4) and (3.6) of Ref.~\cite{sto-therm-NC}.  In the present case, the magnetization $\mv$ plays the role of state variable, and the magnetic field $\Hv$ plays the role of control parameter.  Note that both $\mv$ and $\Hv$ are odd under time-reversal, whereas the spin current vector $\Pv$ is even under time-reversal.  (The fact that $\Pv$ is even under time-reversal can be easily seen via dimensional analysis of Eq.~(\ref{Hv-Hv-f-LLG}).) Hence Eqs.~(3.4) and (3.6) of Ref.~\cite{sto-therm-NC} become  
\begin{subequations}
\label{DB-condition-2}
\ba
 B^{ij}(-\mv, -\Hv)  
&=& B^{ij}(\mv, \Hv ),
\label{DB-condition-B} \\
 Q^{ij}(-\mv, -\Hv)  
&=& - Q^{ij}(\mv, \Hv), \\
 U(-\mv, - \Hv) &=& U(\mv, \Hv), 
 \label{DB-condition-U-1}\\ 
 \int_{\mv}e^{-U(\mv, \Hv)} &=& 1,
\label{normalization-U}\\
 - \varphi_i ( - \mv, \Pv) &=& \varphi_i (\mv, \Pv). 
 \label{DB-condition-phi}
\ea
\end{subequations}
Equations (\ref{DB-condition-B})-(\ref{normalization-U}) are precisely the conditions of detailed balance for systems without spin torque, which were given in Eqs.~(2.29) of Ref.~\cite{LLG-1}.  The new condition (\ref{DB-condition-phi}) can be easily verified using the definition (\ref{varphi-def}).



Let us use $P_{\Hv, \Pv} (\mv_1  | \mv ; dt)$ to denote the probability density function that the system transits to state $\mv_1 = \mv + d \mv$ at time $t+dt$, given that it is in state $\mv$ at time $t$.  Note that the subscripts ${\Hv, \Pv}$ describes the values of magnetic field and spin-current in the dynamics.   {Using Eqs.~(\ref{DB-condition-2}), we may derive the following conditions: }
\ba
 \log  \frac{ P_{\Hv, \Pv} (\mv_1  | \mv ; dt)  }
{P_{-\Hv, \Pv}( -\mv | - \mv_1 ;dt)} 
&=&  \beta \Hv \cdot d \mv  + \beta \mv \times \Pv \circ d \mv
\nonumber
\\
&+&  2\, \gamma_0 \, \mv \cdot \Pv\, dt,
\label{dS_ev-def}
\ea
where $\circ$ is the product in Stratonovich's sense. This relation can be understood as a special case of Eqs.~(3.73) of Ref.~\cite{sto-therm-NC}.  Because of the time-reversal symmetry of the underlying microscopic dynamics, the r.h.s. of Eq.~(\ref{dS_ev-def}) should be understood as the change of environmental entropy as the system transits from $\mv$ to $\mv_1$ in the forward dynamics characterized by $\Hv$ and $\Pv$:
\ba
d \mathscr S_{\rm env} \equiv  \beta \Hv \cdot d \mv
  + \beta \mv \times \Pv \circ d \mv
+  2\, \gamma_0 \, \mv \cdot \Pv\, dt. 
\nonumber\\
\label{dS-env-def}
\ea 
In the absence of spin torque,  Eq.~(\ref{dS_ev-def}) reduces to Eq.~(2.31) of Ref.~\cite{LLG-1}.





\section{Stochastic thermodynamics }
\label{sec:ST}
A general theory of stochastic thermodynamics was developed for non-conservative Langevin  dynamics (\ref{Langevin-def-m-nonc}) in Ref.~\cite{sto-therm-NC}.  This theory may be directly applied to sLLS equation (\ref{Langevin-def-m-vec-NC}).




The fluctuating internal energy is defined the same as in the absence of spin torque (Eq.~(3.1) of Ref.~\cite{LLG-1}):
\be
\mathscr E(\mv, \Hv) \equiv - \mv \cdot \Hv,
\ee
such that  the equilibrium state (\ref{p_eq-U-H}) takes the usual form of Gibbs-Boltzmann distribution:
\ba
p^{\rm eq}(\mv, \Hv) 
= e^{\beta F(\Hv) - \beta \mathscr E(\mv, \Hv)}. 
 \label{p_eq-U-H-1}
\label{p_EQ-2}
\ea

The work and heat at trajectory level are defined as
\begin{subequations}
\label{dW-dQ-def}
\ba
\dbar \mathscr W &\equiv&
  - \mv \cdot d \Hv + \mv \times \Pv \circ d \mv, 
  \label{dW-def}\\
\dbar \mathscr Q &\equiv&
  - ( \Hv +  \mv \times \Pv) \circ d \mv.
  \label{dQ-def}
\ea
\end{subequations}
It is easy to see that the first law of thermodynamics holds at trajectory level:
\ba
d \mathcal H = d\mathscr W + d \mathscr Q. 
\ea

To appreciate the physical meanings of work defined above, we consider the special case of vanishing damping coefficient $\eta = 0$, the sLLS equation (\ref{Langevin-def-m-vec-NC}) reduces to 
\ba
{d \mv} = -  \gamma_0 \mv \times ( \Hv + \mv \times \Pv) \, dt. 
\label{Langevin-eta-0}
\ea
Such a case cane be obtained theoretically by decoupling the system from the heat bath. Since there is no heat, the energy change is entirely due to work. Hence we expect
\ba
\dbar \mathscr W = d \mathscr E(\mv, \Hv) 
= - \mv \cdot d \Hv - \Hv \cdot d \mv.
\label{work-eta-0}
\ea
But according to Eq.~(\ref{Langevin-eta-0}), $ d\mv \cdot(\Hv + \mv \times \Pv) = 0$, and hence Eq.~(\ref{work-eta-0}) is equivalent to Eq.~(\ref{dW-def}).  In other words, Eq.~(\ref{dW-def}) is the correct definition of work at least for the special case $\eta = 0$.   

Using Eq.~(\ref{Langevin-def-m-vec-NC}), we may rewrite the definition of heat Eq.~(\ref{dQ-def}) into:
\ba
\dbar \mathscr Q &=& ( \Hv +  \mv \times \Pv) 
 \circ \big[ \eta \, 
\mv \times (\mv \times ( \Hv +  \mv \times \Pv) )\, dt 
\nonumber\\
&+&   2 T \eta \,\mv \,dt  
-  {\sqrt{2 \eta T}} \mv \times d{\boldsymbol W} \big],
\label{dQ-2}
\ea
which vanishes identically in the limit of vanishing damping coefficient.  Equation (\ref{dQ-2}) may be understood as the work done by the heat bath.  This is of course consistent with the common understanding of heat in stochastic thermodynamics~\cite{Sekimoto-book}. 

Using Eq.~(\ref{dQ-def}), we may rewrite Eq.~(\ref{dS-env-def}) as
\ba
d \mathscr S_{\rm env} &=& - \beta \dbar \mathscr Q 
+ d \mathscr S_{\rm P}  . 
\label{dS-env-def-2}
\ea
 {Whereas $- \beta \dbar \mathscr Q$ 
 is easily understood as the change of bath entropy, the interpretation of the second term in the r.h.s. of Eq.~(\ref{dS-env-def-2}) is very subtle.  It persists even in the limit of vanishing damping, and hence cannot be understood as entropy production due to dissipation.  As discussed in Sec.~III.H of Ref.~\cite{LLG-1}, it should be understood as the entropy being pumped out of the system by the non-conservative force.  In the present case, we believe that it is a non-dissipative entropy transfer from the micro-magnet to the spin current (or the other way around). } Following Sec.~III of Ref.~\cite{sto-therm-NC}, we shall call it {\em the pumped entropy} and denote it as
 \ba
d \mathscr S_{\rm P} = 2\, \gamma_0 \, \mv \cdot \Pv\, dt,
\label{dS_ag-def-1}
\ea
Equation (\ref{dS-env-def-2}) can then be rewritten as
\ba
d \mathscr S_{\rm env} &=& - \beta \dbar \mathscr Q 
+ d \mathscr S_{\rm P} . 
\label{dS-env-def-3}
\ea




The  entropy production at the ensemble level is the sum of change of system entropy and that of the environmental entropy:
\ba
 {d S^{\rm tot}} = {dS} + {dS_{\rm env}}
 = dS - \beta \dbar Q + d S_{\rm P},
\label{dS_tot-1-0} 
\ea
 where $dS$ is the differential of Gibbs-Shannon entropy:
\ba
S[p] = - \int_\mv p(\mv, t) \log p(\mv,t),
\ea
whilst ${dS_{\rm env}}$ is the ensemble average of Eq.~(\ref{dS-env-def-2}).  After some tedious calculation, we obtain  (c.f. Eq.~(3.78) of Ref.~\cite{sto-therm-NC}):
\ba
 \frac {d S^{\rm tot}}{dt} 
&=&T \eta \,  \Big\langle m^2 \big( 
\boldsymbol \nabla \log p - \beta \Hv - \beta \mv \times \Pv \big)^2
\nonumber\\
&  -& \big( \mv \cdot \big( 
\boldsymbol \nabla \log p 
- \beta \Hv - \beta \mv \times \Pv \big) \big)^2 \Big\rangle,
 \label{dS^tot-NC}
  \quad
\ea
where $\langle \, \cdot \, \rangle$ means average over $p(\mv,t)$.  The entropy production is therefore always non-negative, and reduces to Eq.~(3.20) of Ref.~\cite{LLG-1} in the absence of spin torque.  Using Eqs.~(\ref{j_R-IR-def}), it is also easy to verify that the entropy production can be rewritten into 
\ba
 \frac {d S^{\rm tot}}{dt} 
= \int_\mv p^{-1}\, j_{\rm IR}^i B_{ij} j_{\rm IR}^j,
 \label{dS^tot-NC-j-R}
\ea
where  $B_{ij}$ is the generalized inverse matrix of $B^{ij}$. Note that ${d S^{\rm tot}}/{dt} $ depends only on the irreversible probability current $j^k_{\rm IR}$ but not on the reversible current $j^k_{\rm R}$

Because $\mv \times \Pv$ cannot be written as the gradient of any function of $\mv$, it is easy to see that there exists no function $p(\mv)$ such that Eq.~(\ref{dS^tot-NC}) vanishes identically.  For fixed $\Hv$ and $\Pv$, the system converges to a non-equilibrium steady state with positive entropy production rate. 

\section{Fluctuation theorem}
\label{sec:FT}
We consider a {\em forward process} where the system starts at $t = 0$ from the initial equilibrium state $p^{\rm eq}(\mv; \Hv_0)$ as defined in Eq.~(\ref{p_eq-U-H-1}), whereas $\{ \Hv, \Pv \}$ evolve according to the {\em forward protocol}  $\{ \Hv_t, \Pv_t\}$, until $t = \tau$, when the process stops.  We define the {\em backward process} such that the system starts at $t = 0$ from the initial equilibrium state $p^{\rm eq}(\mv; - \Hv_\tau)$ and 
whereas $\{ \Hv, \Pv \}$ evolve according to the {\em backward protocol}  $\{ - \Hv_{\tau -t}, \Pv_{\tau - t}\}$, until $t = \tau$, when the process stops.   Note that both the forward process and the backward process take place in the time interval $[0, \tau]$.  Note also that only the magnetic field $\Hv$ changes sign when we transform the forward process to the backward process.  In general, the system is not in equilibrium either at the end of the process or at the end of the backward process. 

Consider a {\em forward trajectory}:
\ba
 \bm \gamma &=& \{\mv(t), \,\, t \in [0, \tau] \},
 \ea
 we define  its  {\em backward trajectory} as
 \ba
\hat {\bm \gamma} &=&  \{ - \mv(\tau - t),\,\,  t \in [0, \tau] \}. 
\ea 
Let  $\mathscr W_{\rm F} [ \bm \gamma], \mathscr Q_{\rm F} [ \bm \gamma],\mathscr S_{\rm P,F}  [{\bm \gamma}] $ ($\mathscr W_{\rm B} [\hat {\bm \gamma}],\mathscr Q_{\rm B} [\hat {\bm \gamma}],\mathscr S_{\rm P,B}  [\hat{\bm \gamma}]   $) be the integrated work, heat, and pumped entropy along the forward (backward) $\bm\gamma$ ($\hat {\bm \gamma}$) in the forward (backward) process, which can be readily obtained by integrating the differential work and heat that are defined in Eqs.~(\ref{dW-dQ-def}).  We easily find the following symmetry:
\ba
\mathscr W_{\rm F} [ \bm \gamma] &=& 
- \mathscr W_{\rm B} [\hat {\bm \gamma}] 
=   \int_{\bm \gamma} ( - \mv \cdot d \Hv
 + \mv \times \Pv \circ d \mv),
 \nonumber\\
\label{W-F-decomp-1}\\
\mathscr Q_{\rm F} [ \bm \gamma] &=& 
- \mathscr Q_{\rm B} [\hat {\bm \gamma}] 
=  - \int_{\bm \gamma} ( \Hv +  \mv \times \Pv) \circ d \mv,
\label{Q-F-decomp-1}\\
\mathscr S_{\rm P,F}  [{\bm \gamma}]  
&=& - \mathscr S_{\rm P,B}  [\hat{\bm \gamma}]  
=  \int_{\bm \gamma} 2\, \gamma_0 \mv \cdot \Pv\, dt. 
\label{st-pumped-entropy}
\ea
Taking the sum of Eqs.~(\ref{W-F-decomp-1}) and (\ref{Q-F-decomp-1}) we obtain the integrated first law:
\ba
\Delta \mathscr E = \mathscr W_{\rm F} [\bm \gamma] 
+ \mathscr Q_{\rm F} [\bm \gamma],
 \label{1st-law-int}
\ea
where $\Delta \mathscr E $  as the total change of the energy along $\bm \gamma$: 
\ba
\Delta \mathscr E  \equiv \mathscr E(\mv(\tau), \Hv_\tau)
- \mathscr E(\mv(0), \Hv_0), 
\label{Delta-E-def}
\ea

We further introduce ${\bm\gamma}_0  \equiv \mv(0)$ and $\hat {\bm\gamma}_0 \equiv - \mv(\tau)$ to denote the initial state of $ \bm\gamma, \hat {\bm \gamma}$, respectively.  These notations (boldface) should be carefully distinguished from $\gamma_0$, the {\em gyromagnetic ratio}, appearing in Eqs.~(\ref{Langevin-def-m-vec}) and (\ref{Langevin-def-m-vec-NC}).  We can construct the pdfs of trajectories both for the forward process and for the backward process,  using the definition of conditional probability:
\begin{subequations}
 \label{p_FB-gamma}
\ba
p_{\rm F}[ \bm \gamma] &=& 
p_{\rm F}[\bm \gamma | {\bm\gamma}_0] \,
 p^{\rm eq}(\mv(0); \Hv_0),
 \label{p_F-gamma}\\
 p_{\rm B}[\hat {\bm \gamma}] &=& 
p_{\rm B}[\hat {\bm \gamma} |\hat  {\bm\gamma}_0]  \,
p^{\rm eq}(- \mv(\tau); - \Hv_\tau),
 \label{p_B-gamma}
\ea
\end{subequations}
where $p_{\rm F}[\bm \gamma | {\bm\gamma}_0], p_{\rm B}[\hat {\bm \gamma} |\hat  {\bm\gamma}_0]$ are the conditional pdf of trajectories of the forward (backward) processes given their initial states. 

Because of the Markov property, $p_{\rm F}[\bm \gamma  |{\bm\gamma}_0]$ and $ p_{\rm B}[\hat {\bm \gamma} |\hat  {\bm\gamma}_0] $ can be calculated using the time-slicing method.   Further using Eq.~(\ref{dS_ev-def}) for each pair of time-slices, we find
\ba
\log  \frac{ p_{\rm F}[\bm \gamma | {\bm\gamma}_0] }  
{ p_{\rm B}[\hat {\bm \gamma} |\hat  {\bm\gamma}_0] } 
&=& \int_{\bm \gamma}  \left( 
d\mathscr S_{\rm B} + d\mathscr S_{\rm P} \right)
\nonumber\\
&=& {- \beta \mathscr Q_{\rm F}[ \bm \gamma]}
+  \mathscr S_{\rm P,F}[\bm \gamma], 
\label{p-gamma-cond-ratio}
\ea
where $\mathscr Q_{\rm F}[ \bm \gamma]$ is the total heat absorbed by the system along the trajectory $\bm \gamma$ in the forward process. 

Let us define: 
\ba
\Sigma_{\rm F}[ \bm \gamma] &\equiv&
  \log \frac{p_{\rm F}[ \bm \gamma]}{p_{\rm B}[\hat {\bm \gamma}]} .
  \ea
Using Eqs.~(\ref{p_FB-gamma}) and (\ref{p-gamma-cond-ratio}), we obtain:
  \ba
\Sigma_{\rm F}[ \bm \gamma] 
&=&  \log \frac{ p^{\rm eq}(\mv(0); \Hv_0)}
{p^{\rm eq}(- \mv(\tau); - \Hv_\tau)}
-  \beta \mathscr Q_{\rm F}[ \bm \gamma]
+  \mathscr S_{\rm P,F}[\bm \gamma].
\nonumber\\
\label{p-gamma-ratio-1} 
\ea
Recalling the symmetry: ${p^{\rm eq}(- \mv; - \Hv)}
= {p^{\rm eq}( \mv;  \Hv)}$, if the protocol is such that the final state  of the forward process is the equilibrium state $p^{\rm eq}(\mv; \Hv_\tau)$, we may also write Eq.~(\ref{p-gamma-ratio-1}) into 
\ba
\Sigma_{\rm F}[ \bm \gamma] = 
- \log \frac{ p(\mv(\tau), \tau)}{p(\mv(0), 0)}
- \beta \mathscr Q_{\rm F} [ \bm \gamma],
\label{Sigma-gamma-1}
\ea
which is {\em the stochastic entropy production}~\cite{Seifert-2005} along the trajectory $\bm \gamma$ in the forward process.  If the system is not in the NESS at the end of the forward process, however, the physical meaning of $\Sigma_{\rm F}[ \bm \gamma] $ is more subtle.  


Further taking advantage of Eq.~(\ref{p_EQ-2}) as well as the first law (\ref{1st-law-int}), we may rewrite Eq.~(\ref{p-gamma-ratio-1}) into:
\ba
\Sigma_{\rm F}[ \bm \gamma] = 
\log \frac{p_{\rm F}[ \bm \gamma]}
{p_{\rm B}[\hat {\bm \gamma}]} 
=   \mathscr S_{\rm P}[\bm \gamma]
+  \mathscr W_{\rm F} [\gamma]  
 - \Delta F,
\label{p-gamma-ratio-1-1} \quad
\ea
where $\Delta F$ is defined as
\ba
\Delta F \equiv F (\Hv_\tau)  - F(\Hv_0). \label{Delta-F-def}
\ea

Using Eq.~(\ref{W-F-decomp-1}), we may rewrite Eq.~(\ref{p-gamma-ratio-1-1}) into
\ba
\log \frac{p_{\rm F}[ \bm \gamma]}{p_{\rm B}[\hat {\bm \gamma}]} 
&=&  \beta  \left( \mathscr W_{\rm F} [ \bm \gamma] - \Delta F \right)
+  \mathscr S_{\rm P,F}[\bm \gamma]
\nonumber\\
&=&  - \beta \left(  \mathscr W_{\rm B} [\hat {\bm \gamma}]
+  \Delta F \right)
-  \mathscr S_{\rm P,B}[\hat{\bm \gamma}].  
\label{DFT-1}
\ea

\begin{figure}[t!]
    \centering
    \includegraphics[width=3.4in]{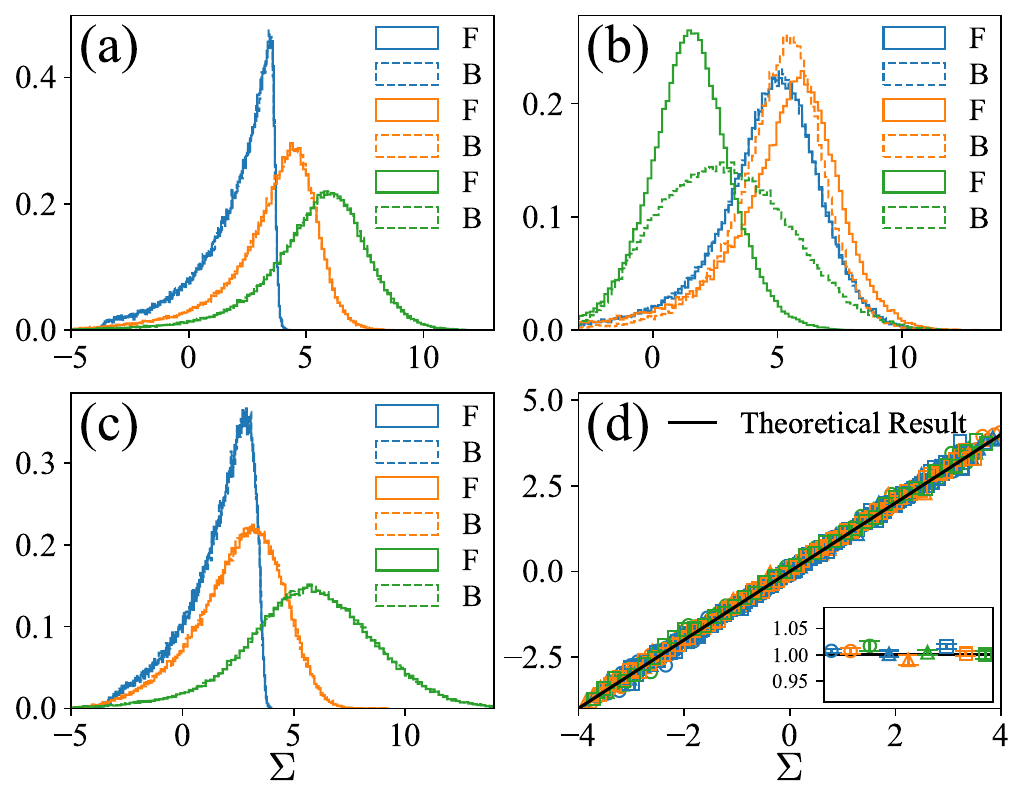}
    \vspace{-5mm}
   \caption{Numerical verification of fluctuation theorem (\ref{st-total-ft}).  The dynamic protocol is shown in Eq.~(\ref{protocol-1}) and Table \ref{tab::ft/ft_torque_combined_TR}.
     (a), (b), (c): Histograms of the entropy production (\ref{Sigma-gamma-def}).       In all legends F, B mean forward and backward respectively.  (d): Verification of FT (\ref{st-total-ft}), where the vertical axis is $\log p_{\rm F} (\Sigma)/p_{\rm B}(-\Sigma)$. The black straight-line is the FT (\ref{st-total-ft}). 
    Circles, triangles, and squares are respectively data from panels (a), (b), (c).  Inset: The fitting slopes and error bars for each process. }
    \label{fig::ft/ft_torque_combined_TR}
   \vspace{-3mm}
\end{figure}
%

We can now define the pdfs of the integrated work both for the forward process and for the backward process:
\ba
p_{\rm F}( \mathscr W,\mathscr S_{\rm P}) 
&\equiv& \int \!\! D {\bm \gamma} \, 
\delta \left( \mathscr W - \mathscr W_{\rm F}[\bm \gamma] \right)
\delta \left( \mathscr S_{\rm P} - \mathscr S_{\rm P,F}[\bm \gamma]\right)
 \, p_{\rm F}[\bm \gamma], 
 \nonumber\\
p_{\rm B}( \mathscr W, \mathscr S_{\rm P})
 &\equiv& \int  \!\! D {\bm \gamma} \, 
\delta \left( \mathscr W - \mathscr W_{\rm B}[\bm \gamma] \right)
\delta \left( \mathscr S_{\rm P} - \mathscr S_{\rm P,B}[\bm \gamma]\right)
 \, p_{\rm B}[\bm \gamma]. 
\nonumber
\ea
Taking advantage of Eq.~(\ref{DFT-1}), and using standard methods of stochastic thermodynamics, we can establish the following {\em fluctuation theorem} for $\mathscr W$ and $\mathscr S_{\rm P}$:
\be
p_{\rm F}(\mathscr W, \mathscr S_{\rm P}) \,
e^{-\mathscr S_{\rm P} - \beta \mathscr W + \beta \Delta F} 
= p_{\rm B}(- \mathscr W, -\mathscr S_{\rm P}). 
\label{FT-W-SP}
\ee
We can also obtain the generalized Jarzynski equality:
\be
\left\langle e^{- \beta \mathscr  W -  \mathscr S_{\rm P}} \right\rangle 
 = e^{-\beta \Delta F}.
\ee

We define the following functional of a trajectory $\bm \gamma$:
\ba
\Sigma [ \bm \gamma] \equiv 
\mathscr S_{\rm P} [ \bm \gamma] + \beta \mathscr W [ \bm \gamma] 
- \beta \Delta F,
\label{Sigma-gamma-def}
\ea
which may be understood as the stochastic entropy production along $\bm \gamma$, if the final state of the system is an equilibrium state.  We can then use Eq.~(\ref{FT-W-SP}) to establish the following fluctuation theorem:
\ba
p_{\rm F}( \Sigma)\,e^{- \Sigma } = p_{\rm B}(-\Sigma). 
\label{st-total-ft}
\ea

\begin{figure}[t!]
    \centering
   \includegraphics[width=3.4in]{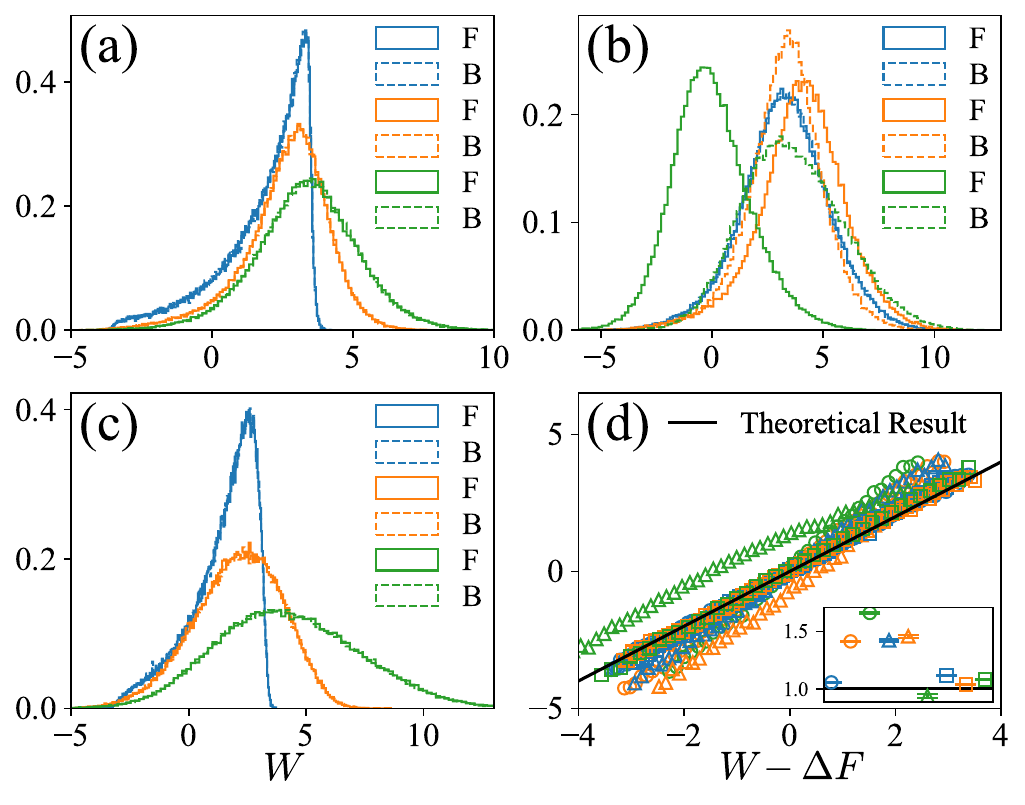}
    \caption{ Verification that work $\mathscr{W}$ does not satisfy the usual fluctuation theorem. Vertical axis is $\log p_{\rm F} (\mathscr{W})/p_{\rm B}(-\mathscr{W})$. }
    \label{fig::ft/ft_s_torque_combined_TR-g0}
\end{figure}


Recall that from the joint pdf $p_{\rm F}(\mathscr W, \mathscr S_{\rm P})$ we can define the conditional pdfs for $W$ and for $S_{\rm P}$ respectively:
\ba
p_{\rm F}(\mathscr W| \mathscr S_{\rm P}) &=& 
\frac{p_{\rm F}(\mathscr W,\mathscr S_{\rm P})}{p_{\rm F}(\mathscr S_{\rm P})}, 
\\
p_{\rm F}(\mathscr S_{\rm P} | \mathscr W) &=& 
\frac{p_{\rm F}(\mathscr W, \mathscr S_{\rm P})}{p_{\rm F}(\mathscr W)}. 
\ea
We can also define the conditional averages:
\ba
\left\langle e^{- \mathscr S_{\rm P}} \right\rangle_{\mathscr W} 
&\equiv& \int_{\mathscr S_{\rm P}} e^{-\mathscr S_{\rm P}}
 p_{\rm F}(\mathscr W|\mathscr S_{\rm P}),
\\
\left\langle e^{- \mathscr W} \right\rangle_{\mathscr S_{\rm P}}
&\equiv& \int_{ \mathscr W} e^{-\beta \mathscr W} 
 p_{\rm F}(\mathscr S_{\rm P} | \mathscr W).
\ea
These two conditional averages are respectively functions of $W$ and of $S_{\rm P}$.  
From Eq.~(\ref{FT-W-SP}) we can then derive the following fluctuation theorems for $\mathscr S_{\rm P}$ and for $\mathscr W$:
\ba
p_{\rm F}(\mathscr W) e^{- \beta \mathscr W}  \left\langle
 e^{- \mathscr S_{\rm P}} \right\rangle_{\mathscr W} 
&=& p_{\rm B}(- \mathscr W) \,e^{-\beta \Delta F}, 
\quad\quad \label{FT-W-cond}\\
p_{\rm F}(\mathscr S_{\rm P}) e^{-\mathscr S_{\rm P}}  \left\langle 
e^{\beta ( \Delta F - \mathscr W)}  \right\rangle_{\mathscr S_{\rm P}} 
&=& p_{\rm B}(- \mathscr S_{\rm P}). 
\label{FT-SP-cond}
\ea
Note that Eq.~(\ref{FT-W-cond}) reduces to the usual Crooks FT:
\ba
p_{\rm F}(\mathscr W) e^{- \beta \mathscr W} 
&=& p_{\rm B}(- \mathscr W) \,e^{-\beta \Delta F}, 
\label{CFT}
\ea
if entropy pumping is absent.

\subsection{Verification of fluctuation theorem of $\Delta S^{\rm tot}$}

\begin{table}[th!]
    \centering
    \renewcommand{\arraystretch}{1.6}
    \tabcolsep=7pt
    \begin{tabular}{|c|c|c|c|c|}
        \hline
        process & color & $\tau$ & $\boldsymbol{H}_\tau$ & $\boldsymbol{P}_{\tau/2}$ \\ \hline
        \multirow{3}{*}{(a)} & blue & 0.1 & \multirow{3}{*}{(0,0,-1)} & \multirow{3}{*}{(0,0,1)} \\ \cline{2-3}
        & orange & 1 &  &  \\ \cline{2-3}
        & green & 2 &  &  \\ \hline
        \multirow{3}{*}{(b)} & blue & \multirow{3}{*}{1} & (0,0,-1) & \multirow{3}{*}{(0,1,1)} \\ \cline{2-2} \cline{4-4}
        & orange &  & (0,-1,-1) &  \\ \cline{2-2} \cline{4-4}
        & green &  & (0,1,1) &  \\ \hline
        \multirow{3}{*}{(c)} & blue & \multirow{3}{*}{1} & \multirow{3}{*}{(0,0,-1)} & (0,0,0.2) \\ \cline{2-2} \cline{5-5} 
        & orange &  &  & (0,1,0) \\ \cline{2-2} \cline{5-5} 
        & green &  &  & (0,2,0) \\ \hline
    \end{tabular}
    \caption{Parameters used in the simulation study. All process has the same parameter $\boldsymbol{H}_0 = (0,0,1), \eta=0.5, \gamma_0=1, T=1$. }
    \label{tab::ft/ft_torque_combined_TR}%
\end{table}

We verify the Fluctuation Theorem Eq.~(\ref{st-total-ft}) by numerical simulation of the sLLS equation (\ref{Langevin-def-m-vec-NC}).  Details of the simulation method was discussed in Appendix A of Ref.~\cite{LLG-1}.   The entropy production of trajectory $\bm \gamma$ is calculated using Eqs.~(\ref{Sigma-gamma-def}), (\ref{W-F-decomp-1}), and (\ref{st-pumped-entropy}).  We simulate the following protocols for the forward process:
\begin{subequations}
\ba
\Hv_t &=& \Hv_0 + \frac{t}{\tau} \left( \Hv_\tau  - \Hv_0\right),\\
\Pv_t &=& \left\{ \begin{array}{ll}
\frac{2 t}{\tau} \Pv_{\tau/2}, &  0 \leq t \leq \tau/2;
\vspace{2mm}\\
\frac{\tau - 2 t}{\tau} \Pv_{\tau/2}, \quad\quad &  \tau/2 \leq t \leq \tau.
\end{array}
\right.
\ea 
\label{protocol-1}
\end{subequations}
Note that the spin current vector $\Pv$ vanishes identically both at the beginning and at the end of the process.  Furthermore, the initial state of the forward (backward) process is the equilibrium state corresponding to the magnetic field $\Hv_0$ ($\Hv_\tau$), as we discussed above.  The duration  $\tau$ of the process, the damping coefficient $\eta$, the initial and final fields $\Hv_0, \Hv_\tau$, as well as the peak value of the spin current vector $\Pv_{\tau/2}$ are systemically varied, as shown in Table \ref{tab::ft/ft_torque_combined_TR}.   The verification of the FT (\ref{st-total-ft}). is shown in Fig.\ref{fig::ft/ft_torque_combined_TR}. As one can see there, all data agree well with the theoretical prediction, which is shown as the solid straight-line in (d), (e), and (f).

Using the same simulation data, we may also verify that the usual form of Crooks fluctuation theorem (\ref{CFT}) is not satisfied, due to the existence of entropy pumping.  The results are shown in Fig.~\ref{fig::ft/ft_s_torque_combined_TR-g0}. 
%

\section{Conclusion}
\label{sec:conclusion}

In this work, we have demonstrated that interaction with a spin-current torque leads to  important change of the stochastic thermodynamics of micromagnetics.  The new effect of entropy pumping may be used to design novel steady-state information engine that extract useful works from heat bath or reduce the entropy of spin-polarized current.  The physical properties of these systems will be studied in future publications.  




The authors acknowledge support from NSFC via grant 11674217(X.X.), as well as Shanghai Municipal Science and Technology Major Project (Grant No.2019SHZDZX01).

\end{document}